# A Toolkit for Scalable Spreadsheet Visualization

*Markus Clermont*
*Software Quality Research Laboratory*
*Dept. of Computer Science and Information Systems*
*University of Limerick*
*IRELAND*
*E-Mail: markus.clermont@ul.ie*

## ABSTRACT

*This paper presents a toolkit for spreadsheet visualization based on logical areas, semantic classes and data modules. Logical areas, semantic classes and data modules are abstract representations of spreadsheet programs that are meant to reduce the auditing and comprehension effort, especially for large and regular spreadsheets.*

*The toolkit is integrated as a plug-in in the Gnumeric spreadsheet system for Linux. It can process large, industry scale spreadsheet programs in reasonable time and is tightly integrated with its host spreadsheet system.*

*Users can generate hierarchical and graph-based representations of their spreadsheets. This allows them to spot conceptual similarities in different regions of the spreadsheet, that would otherwise not fit on a screen. As it is assumed that the learning effort for effective use of such a tool should be kept low, we aim for intuitive handling of most of the tool's functions.*

## 1   INTRODUCTION

In today's business life many important decisions are based on the results of spreadsheet programs. For a software engineer a spreadsheet program is obviously software and thus should be developed obeying some systematic approach and then be carefully tested. For the typical spreadsheet user, who is not a software engineer, but rather an expert in the application domain, a spreadsheet program is not considered as software but as a tool for performing calculations and formatting their results. Spreadsheets are often considered as a word-processor for numbers, and not as the highly complex data flow program that they really are.

Another issue that we often observe is overconfidence. Even if the spreadsheet users are aware that their spreadsheet is complex and the occurrence of errors is likely, they do not believe that there could possibly be an error in any spreadsheet that they have written. In field audits we have observed a case were the author of a spreadsheet denied the presence of errors in his spreadsheets even when they were pointed out to him, although they had a severe impact on the spreadsheets' results.

So, despite the importance of spreadsheets there is still only little effort made to assure the correctness of spreadsheet programs. Although there are already a couple of possible methodologies to either inforce a systematic development of spreadsheets, according to software engineering principles (see [Chadwick 1999, Isakowitz 1995, Ronen 1989, Knight 2000] or to reduce the error rate of already existing spreadsheets, by testing (see [Ayalew 2001, Rothermel 2000, Burnett 2000]) or auditing (see [Butler 2000, Sajaniemi 2000], and other commercially distributed tools), they are still not widely accepted. One reason for the poor acceptance of approaches that require a systematic development of spreadsheet programs is the nature of the spreadsheet itself. As they are meant to be a modelling tool for end users (see Nardi 1990, Brown 1987) it seems exaggerated to force the users into a distinct design phase before implementing the actual spreadsheet. Additionally, spreadsheet programmers are end users and would need extra training in order to apply a systematic approach.

Unfortunately, the lack of systematic development is not counter-balanced by rigorous checking of the finished spreadsheet. As it was found out earlier by Panko (see [Panko 1998]) checking a spreadsheet is a time consuming and expensive task. Thus, it should be limited to the crucial parts of spreadsheets that are most likely to be subject to errors. However, it is not trivial to identify these parts in a quick and efficient way. There are some methodologies that operate on user assesment of the risk of an error in a certain region of the spreadsheet (see [Butler 2000]), but they are subject to the auditors attitude and might not map to the actual erroneous areas of spreadsheet programs. Although there are approaches to direct the auditors' attention to certain parts of the spreadsheet, mainly by means of visualizing the distribution of equal formulas throughout the spreadsheet, the approaches existing so far work fine for small examples, but are not efficient to handle large, 'industry-sized' spreadsheets of some 10000 cells or more.

In this paper we will introduce a toolkit that supports auditors to identify the hot-spots in spreadsheet programs based on the concepts of semantic classes and data modules (see [Mittermeir 2002, Clermont 2003.2, Clermont 2003.3]). Both, conceptually and technically it is a further development of an auditing toolkit that was used in a case study also presented at this conference (see [Clermont 2002]). Therefore, we will briefly summarize these concepts in the next section. Section 3 will outline the usage of our prototype and in section 5 we will demonstrate its capabilities by means of an example.

## 2  SEMANTIC CLASSES AND DATA MODULES

Our auditing toolkit is mainly based on two orthogonal methodologies to identify related cells in a spreadsheet. The first methodology tries to identify semantic classes, i.e. blocks of similar cells, where the criterion for similarity is based on the structure of a cell's formula. The second concept, data modules, is based on identifying blocks of cells, that are tightly linked by cell references.

### 2.1  Semantic Classes

As mentioned above, a semantic class can be considered as a set of blocks of similar cells (a more formal definition is given by [Mittermeir 2002]). These blocks, in the rest of the paper called semantic units, have to satisfy certain geometric conditions that can restrict their horizontal and vertical extension as well as the size of gaps in these blocks. In the current toolkit the geometrical conditions have to be supplied by the users by means of three parameters: $d_h$, $d_v$ and $d_{Man}$.

The first two specifiy the maximal size of gaps in the semantic unit, either horizontally ($d_h$) or vertically ($d_v$). Thus, by setting $d_h$ to 1 and $d_v$ to 0 users can require semantic units to consist of horizontally adjacent cells. Setting $d_h$ to 2 and $d_v$ to 0 allows semantic units to consist of horizontally adjacent cells, with gaps spanning at most one cell.

If both $d_h$ and $d_v$ are set to values greater than zero, semantic units are can be of a rectangular shape. As users will sometimes want to restrict the overall size of gaps in these rectangles, the parameter $d_{Man}$ allows to specify the maximum manhattan distance[1] that is spanned by a gap. Setting $d_h$ and $d_v$ both to 2 would result in rectangular semantic units, with horizontal and vertical gaps (see Figure 1). However, in order to forbid the maybe unwanted case shown in Figure 1 in rows 11 to 12, $d_{Man}$ can be set to 1.

---

[1] The Manhattan distance between two cells with the coordinates $(x_1, y_1)$ and $(x_2, y_2)$ is $|x_1-x_2| + |y_1-y_2|$.

In order to group a couple of semantic units into a semantic class, they are required to be similar. Two semantic units are considered similar, if they have an identical geometrical shape and extent, and all the cells on same relative positions in the semantic units are similar. Two cells are considered similar, if their formulas are either

- copy-equivalent, i.e. they are absolutely identical, maybe created by copying and pasting a cell, or
- logical-equivalent, i.e. they differ only in absolute cell references[2] and constant values, or
- structural-equivalent, i.e. they differ in relative[3] and absolute cell references and in constant values. Structural equivalence means there are the same operators in the same order.

Figure 1: Possible shapes of semantic units with $d_h$ and $d_v$ set to 1 and $d_{Man}$ set to 2. The shape in rows 11 and 12 would be impossible if $d_{Man}$ was set to 1.

These similarity criteria comply with the so-called node-equivalence-classes that have been introduced in [Clermont 2003] in the context of logical areas. It is shown that there exists an ordering relation between the degrees of similarities, because two copy-equivalent formulas are also logical equivalent, and two logical equivalent formulas have to be structural equivalent, too. Thus, it is said that structural equivalence is weaker than logical equivalence, and logical equivalence, again, is weaker than copy equivalence.

Thus, a semantic unit is a set of cells that satisfy the imposed geometrical restrictions and a semantic class is a set of semantic units with cells on the same position relative to the semantic units' upper left cell are in the same logical area, i.e. are either copy-, logical- or structural equivalent. Users can control the requested degree of similarity between the semantic units between two further parameters, $eq_{Start}$ and $eq_{Rest}$. Both specify the requested similarity between the cells in semantic units in the same semantic class. $eq_{Start}$ specifies only the similarity for the top-left cells, whereas $eq_{Rest}$ is applied to all the other cells.

---

[2] An absolute cell reference is a cell reference that specifies the coordinates of the referenced cell relative to the upper left corner of the spreadsheet (see [Clermont 2003.2]). In all major spreadsheet systems an absolute cell reference is marked by the $-sign in front of the cell coordinates.
[3] A relative cell reference gives the coordinates of the referenced cell relative to the referencing cell.

This separation was originally introduced to decrease the number of automatically defined semantic classes, by allowing a higher degree of similarity between the starting cells than between the rest of the cells.

### 2.1.1 Auditing Strategies

Semantic classes are meant to generate a compact abstract representation of the spreadsheet. Each semantic class represents a set of similar semantic units, and again each semantic unit represent a set of geometrically and because of the recurring pattern, also conceptually related cells. Hence, a straight-forward, but effective auditing strategy is:

1. Identify the semantic units and semantic classes
2. Check whether semantic units in the same semantic class are distributed according to a certain geometrical pattern on the spreadsheet
3. Find outliers or irregularities in the pattern and add them to the list of hot-spots.
4. Repeat 2 and 3 for all semantic classes
5. Check the hot-spots in detail, using any effective auditing or testing technique.

Obviously, testing or auditing is still necessary, but the amount of cells that has to be checked is efficiently reduced. Furthermore, most of the irregularities are very easy to spot, because semantic units correspond very often to whole rows or columns of the spreadsheet to examine. In contrast to many existing approaches that are considered inefficient, the irregularities can be spotted on a higher level of abstraction and are thus, easier to find.

Besides searching patterns, there are two more effective auditing strategies. The first one relies on the assumption, that errors do very often materialize themselves by means of slight deviations, i.e. a misreference or a constant instead of a cell reference. Thus, comparing the result of the analysis of the spreadsheet with varying similarity criteria effectively points out hot-spots. If the two results are not identical, auditors see themselves confronted with the question: Why are specific semantic units logical equivalent, but not copy-equivalent? Of course, sometimes this is just what the spreadsheet users wanted, but sometimes it is a strong clue for an error.

The third auditing strategy is based on the partly data-flow nature of a spreadsheet program. Therefore, the way a given spreadsheet works can be reconstructed by examining the cell-dependencies. Unfortunately, for large spreadsheet programs, the data-dependency graph (DDG)[4] becomes two complex to be comprehended with reasonable effort. Thus, a more compact, but still meaningful representation has to be offered. We create an abstraction of the DDG, the so-called SRG[5], that is a directed graph, where each semantic unit is represented by a vertex, and there is an edge between two vertices $v_1$ and $v_2$, if any cell in $v_2$ references any cell in $v_1$. A detailed inspection of the SRG can help to give insight into the way the spreadsheet works, and direct the auditors' focus on details that might be hidden in the mass of information that is shown in the DDG.

A more detailed discussion of auditing strategies that rely on semantic classes would be beyond the scope of this paper and is given in ([Clermont 2003.1] and [Clermont 2003.3]).

### 2.2 Data modules

In this section we give only a brief summary of the concept of data modules. For a more detailed discussion, we refer to [Clermont 2003.3]. Spreadsheet programs have some basic characteristics of data flow programs and of graph-reduction programs, too (see [Mittermeir 2004]). Thus, the DDG of a spreadsheet program has an important role for its execution. As the DDG is a directed, acyclic graph, there are some nodes, that are not sources of further edges, i.e. sink nodes.

---

[4] The DDG is a directed, acyclic graph, very edge vertice corresponds to a cell in the spreadsheet. There is an edge between two vertices ($v_1$, $v_2$) if the formula in $v_2$ references the value displayed in $v_1$. Empty cells that are not referenced by any other formula are not represented in the DDG.

[5] SRG is the acronym for set relation graph, because each node in the SRG represents a set of cells.

To grasp the idea, one can assume that a data module is a set of cells that has a distinguished result cell, that is transitively dependent on all cells in the data module. Cells that are outside the data module may only reference its result cell. Broadly speaking, a data module is a subgraph of the DDG, that has only a single sink node, namely its result cell. The result cell of such a data module is either a sink node of the DDG, i.e. a result cell of the spreadsheet program, or a node that is connected to more than one data module. Cells that are not part of a specific data module may reference only its result cell. Obviously, this definition is recursive, but because of the hierarchical organisation of a DDG and its finiteness, this is not a problem.

As the data modules are not a-priori known, we have developed a way to recover them out of the spreadsheet's DDG. The recovery of data modules will start assuming the spreadsheet's result cells to be data modules and adds all cells that are only referenced by one data module to this data module. A cell that transitively contributes to more than one data module is assumed to be the starting point of a new data module and will be treated in the same way.

However, before the DDG can be partitioned into such data modules, the result cells have to be identified. Obviously, not all sink nodes of the DDG have the semantics of a result of the spreadsheet program, e.g. check-sums. In contrast to conventional programming where intermediate results are not displayed and each subroutine has a well defined result, in a spreadsheet each intermediate result is visible to the user and all the other formulas. Sometimes, calculations are deliberately formulated in a more complicated way in order to obtain some desired intermediate results.

Obviously, most of the cells of a spreadsheet program can be considered as well as a computational auxiliary, intermediate result or result cells. For sure it can only be said that cells that are not further referenced by other cells are result cells, because we know that users place them on the spreadsheet with the single purpose to see their contents.If they would not like to see the displayed value, they had not introduced this cell.

Therefore, it seems legitimate to consider DDG sink nodes, i.e. cells, that are not referenced by other cells as result cells, and search those cells, that influence a specific result. As a matter of fact, it is often the case, that the sink nodes in the DDG are not the real results, but check-sums. In this case, the check-sums have to be removed manually, and the remaining DDG is then analyzed.

### 2.2.1 Auditing Strategies

Data modules are particularly useful to identify errors due to misreferences. If a intended cell reference is not part of a formula due to an error, the data module will split up in two different modules. In the opposite case, that a cell reference that should not be part of a formula, might lead to the merge of two unrelated data modules.

Hence, auditors have to watch out for superfluous data modules. Subsequently, the cell where the result of the superfluous data module should have been referenced has to be identified and corrected. The opposite case is more difficult. If an expected data module is not part of the visualization, auditors have to look for the cell where the missing data module is erroneously referenced.

Although fault tracing is more troublesome, the presence of an error can be easily detected. In contrast, certain kinds of errors that are easily discovered by other techniques do not influence the resulting data modules at all. E.g., wrong operators or mis-references to cells in the same data modules, will influence the result of a data module, but not the assignment of cells to a data module, as only the data dependencies are taken into account.

A different auditing strategies makes again use of the fact that we can generate a compressed, but semantically equivalent, representation of the DDG. In the so generated SRG, each data module is a node and there is an edge between data modules, if one references the result cell of the other one. Assuming that original DDG is acyclic, the SRG will be acyclic, too.

The SRG can be used to generate a fish-eye view of the spreadsheets, if we replace one of the data modules by the subgraph of the DDG that it corresponds to. Thus, we can have a very detailed look at a certain part of the spreadsheet, without being bothered by unnecessary details, but still having an eye on the context of the part we are currently examining.

### 2.2.2 Fault Tracing

Fault tracing is a very common problem in spreadsheet programs, as the symptoms of errors often do not occur at the same place as the faults that cause the wrong results. Hence, most testing techniques also involve techniques for fault tracing that are usually based on the calculation of error probabilities for the predecessors of the faulty cell in the spreadsheet programs DDG (see e.g. [Ayalew 2001] or [Rothermel 2000]).

The generation of data modules and the usage of the SRG are powerful helps for fault tracing. If an error is detected in the result cell of a data module, it is not necessary to check all the predecessors in the DDG until the error is found. If the spreadsheet auditor is aware of the data module where the symptom of the error occurred, there are only two possibilities:

1. The error occurred inside the data module where it is detected, or
2. the error occurred in a predecessor module in the SRG.

It is not difficult to decide on which case applies: the spreadsheet auditor has to check the result cells of the predecessor data modules in the SRG. If they are correct, the error is buried in the module where the failure occurred. Else it is assumed that the error is propagated from the erroneous module.

For the first case, the DDG of the data module where the failure occurred has to be checked by one of the techniques that are suggested in [Ayalew 2001] and [Rothermel 2000]. Nevertheless, as a piece of extra information, the auditors are aware that the error must be in the currently examined subgraph of the DDG, and the bug tracing can stop at the module boundaries.

In the second case, the same process is repeated: it has to be checked, whether the fault occurred inside the data module, or in one of its predecessor modules. Depending on the error source, either the module is checked, or the search continues upward in the SRG.

Obviously, also a combination of error sources is possible, as errors can be hidden inside the module as well as in several predecessor modules. Nevertheless, an iteration of several testing and correction phases will finally find all the errors. In this case, the SRG is helpful because it can decrease the number of entities that have to be examined in order to find all the errors.

Next we will introduce the analysis functionality of our auditing toolkit using the concepts of data modules and semantic classes. A brief introduction into the analysis of spreadsheet programs using logical areas is given in [Mittermeir 2000] and [Ayalew 2000].

## 3  A SPREADSHEET AUDITING TOOLKIT

Although Excel is currently the most popular spreadsheet system, Gnumeric was chosen as target system for the prototype of the spreadsheet auditing toolkit. Due to the fact that Gnumeric is subject to the GNU-public license (see [GNU 2003] for the details), the source code is public domain, and can thus be modified and used freely.

In contrast, the Excel source code is not available at all. Visual Basic for Applications (VBA) offers extensive functionality to increase the functionality of Excel. It turned out that a first prototype of the auditing toolkit that was written with VBA had massive performance problems, because in order to assign cells to logical areas, that is a necessary task for building semantic classes later, the abstract syntax trees of the attached formulas have to be compared. As the Excel formula parser is not accessible from VBA, each formula had to be parsed again by a parser that used to be part of the prototype initially.

### 3.1  Advantages of Gnumeric

Although Excel has the advantage to be the most wide-spread spreadsheet system, there are some technical advantages that favored Gnumeric, as the here introduced auditing toolkit was developed merely as a research protoype. In order to make it accessible to a large market, it is necessary to revise the decision about the target-system. However, this discussion is out of scope for this paper.

Gnumeric plug-ins can be developed in any programming language that seems appropriate. They just have to register themselves to the spreadsheet systems by means of a specific plug-in API. As Gnumeric loads plug-ins dynamically into its address space at runtime, plug-ins can access all internal functions and data structures of the spreadsheet system1. Thus, not only the integration of a parser into the auditing toolkit becomes superfluous, but also the parsing of individual cells for comparing abstract syntax trees needs not be done by the add-on, because each formula is parsed as soon as it is entered and the abstract syntax tree is stored in an internal data structure.

Hence, the already stored abstract syntax tree can be used for comparing formulas. The performance of the prototype had been heavily improved by reimplementing it in the Linux and Gnumeric environment, because
1. runtime performance of C is superior to VBA and
2. formulas do not have to be parsed at analysis time.

Further advantages concerning the accessible internal functions of the Gnumeric spreadsheet system have supported the decision. These internal functions, e.g. parsing cell references of the A1 style and converting them to the R1C1 style, managing ranges and evaluating specific cells, saved a lot of development time and eliminated several potential sources of errors.

### 3.2  Usage

The prototype is invoked via an entry in the spreadsheet system's main menu. Once opened, users see a dialog that offers them currently three possibilities:
1. Analysis by Logical Areas
2. Analysis by Semantic Classes
    3. Analysis by Data Modules

As the first issue is already discussed at another location, we will concentrate on the latter two.

### 3.2.1 Using the Prototype for Semantic Class Analysis

In order to analyse a spreadsheet for semantic classes, users have to enter the required parameters, i.e. $d_h$, $d_v$, $d_{Man}$, $eq_{Rest}$ and $eq_{Start}$, in the dialog (see the left dialog in Figure 2). After pressing the analysis button the abstraction of the spreadsheet is generated, using the algorithms discussed in [Clermont 2003.3]. The result is represented in a tree-control, where each entry corresponds to a semantic class. Attached to each semantic class are its member semantic units.

Additionally, a second dialog opens that displays a visualization of the SRG. The SRG is visualized by the graph visualization package LEDA and AGD (see [Ganser 1999]).

In each of the offered visualizations, users can select an entry (i.e., a semantic class or a semantic unit in the tree control, or a node in the SRG), and the corresponding cells in the spreadsheet are selected. This feature can be used for instant coloring of cells, or for the identification of patterns.

### 3.2.2 Using the Prototype for Data Modules Analysis

In order to reconstruct the data modules of a spreadsheet we must, at first hand, identify the result nodes of the spreadsheet. Therefore, users are presented a list of sink nodes in the spreadsheet's DDG as well as a graphical display of the DDG, with sink nodes colored red. However, the later is sometimes no useful help, as DDG with more than 1000 nodes and 10000 edges do occur. The dialoge pictured in Figure 2 on the right hand side shows the results for the analysis of a spreadsheet that has obviously two 'result' cells (see Figure 3, on the left). If the user decides to remove one of the sink nodes from further analysis, maybe because it represents a check-sum, it is removed from the list of sink nodes and replaced by its predecessors in the DDG.

Again, the result of the analysis Is shown in the structure of a tree, with each data module and its associated cells forming one entry in the tree. Users can also examine the SRG (see Figure 3, on the right) and select a data module in the tree view and zoom into it (see the resulting SRG in Figure 4). In this case the displayed SRG is modified, such that the node that corresponds to the data module zoomed into is replaced by the subgraph of the DDG that contains its member cells and their references.

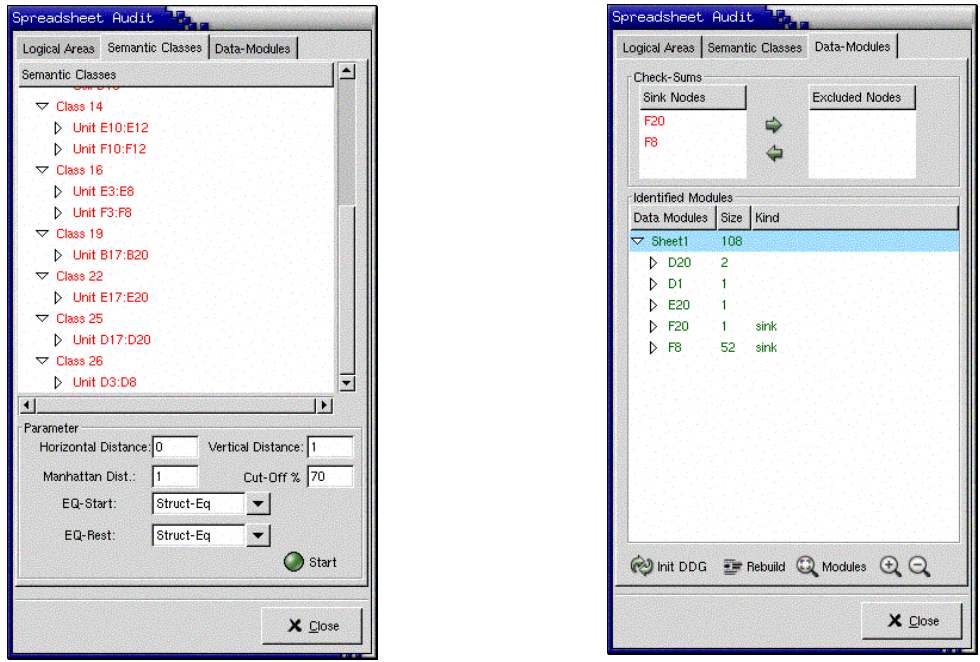

Figure 2: The Semantic-Classes Analysis Dialog and the Data Modules Analysis Dialog. The analyzed spreadsheet is not included in this paper, but is published in [Panko 1997].

## 4   DISCUSSION

The main motivation behind the development of this toolkit was the lack of scalable auditing tools at the moment. The viability of most auditing tools is demonstrated only on small scale laboratory examples and most of them do not hold in the large what they promise in the small. Most of the existing approaches perform very well on small spreadsheets, or by working on one part after the other, but, unfortunately, the effort seems to grow exponentially with the size of the analyzed spreadsheet.

Obviously, as there are different kinds of spreadsheet, different approaches have to be selected for each type of spreadsheets. As a matter of fact, we found out, that semantic classes deliver very good abstractions of large spreadsheets that are usually created by copying, pasting and modifying cells. One of the example spreadsheets that we analyzed consisted of some 1200 formula cells, but boiled down to 23 semantic classes. There are other spreadsheets that do not partition well into semantic classes, usually consisting of a high number of semantic classes with only a few members - but some of them are well suitable for the analysis by data modules, e.g. if they consist of more or less independent parts that are aggregated into the calculation of some business figures.

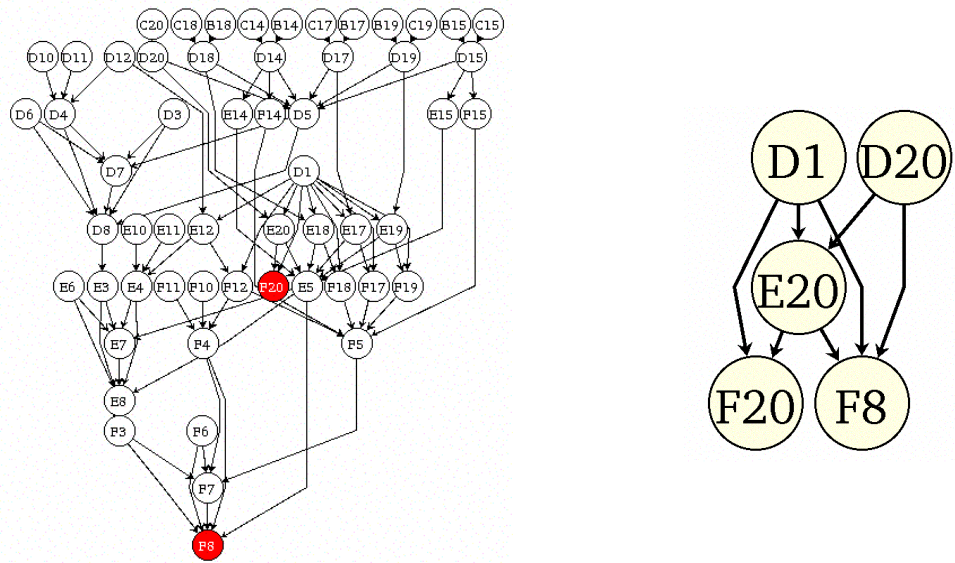

Figure 3: On the left hand side, the DDG of a small example spreadsheet (taken from ...) with 68 nodes and 86 edges. On the right hand side the SRG representing data modules as nodes of the same spreadsheet.

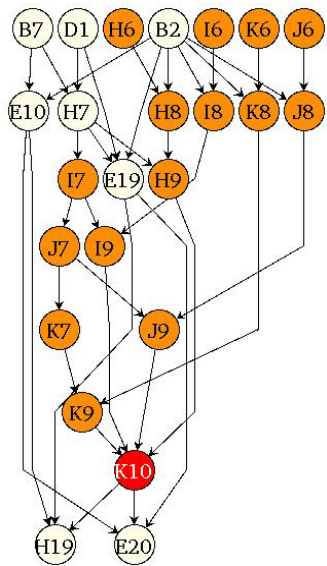

Figure 4: The same SRG as in Figure 3, right hand side, but with an expansion of the node originally labeled K10 .

Each of the so discovered data modules or semantic classes can than be checked independently, using any other test methodologies. For semantic classes, the effort is reduced by checking only one semantic unit in a class - if this one is correct, we have to verify the places where it turns up only. Data modules allow auditors to check one part of the spreadsheet at a time without losing information about the context.

Unfortunately, we cannot make any claim on the correctness of a spreadsheet. Even if the analysis with semantic classes or data modules does not show any irregularities, and every formula is correct, the spreadsheet also depends on the input-values, that are not considered by this approach.

Nevertheless, the here introduced tool offes functionality to easily discover constants in the spreadsheet and in case that an erroneous result is discovered, data modules will support the actual fault tracing.

**5 FURTHER WORK**

Although the here introduced toolkit offers an efficient way to discover irregularities, even in large spreadsheets, there are, of course, many points that can be improved. To start with, the auditing toolkit is a prototype and, thus, the user interface needs still improvement. Additionally, the fact that the auditing toolkit requires a Linux-box running the gnumeric spreadsheet system is a drawback if you try to open to a wide market. Hence, integeration with Excel - and re-implementation of the prototype in a Windows environment - seems necessary.

Additionally, the underlying concepts could be extended in manifold ways. One of the most promising that is currently dealt with in a upcoming master-thesis (see [Hipfl 2004]) includes layout information into the analysis and can automatically identify the parameters for the semantic class analysis. Furthermore, it is no longer necessary to have the same parameters applied to the whole spreadsheet. This approach has already been integrated into the here presented prototype (see [Hipfl 2004] for more details).